# Anisotropy enables advanced light sheet microscopy with opaque lenses

**Mikis Mylonakis [1,2], Evangelos Marakis[1], George J. Tserevelakis[1], Athanasios Zacharopoulos[1], Maira Tampakaki [3,4], Iro Manne [1,5], Angela Pasparaki [6], Nektarios Tavernarakis[4,6], Dimitrios G. Papazoglou[1,2] and Giannis Zacharakis [1,*]**

[1] *Institute of Electronic Structure and Laser, Foundation for Research and Technology-Hellas, N. Plastira 100, Vasilika Vouton, 70013 Heraklion, Crete, Greece*
[2] *Materials Science and Engineering Department, University of Crete, 71003 Heraklion, Greece*
[3] *Institute of Computer Science, Foundation for Research and Technology-Hellas, N. Plastira 100, Vasilika Vouton, 70013 Heraklion, Crete, Greece*
[4] *School of Medicine, University of Crete, 71003 Heraklion, Greece*
[5] Depart of Physics, University of Crete, 70013 Heraklion, Greece
[6] Institute of Molecular Biology and Biotechnology, Foundation for Research and Technology-Hellas, N. Plastira 100, Vasilika Vouton, 70013 Heraklion, Crete, Greece
[7] Department of Biology, University of Crete, 70013 Heraklion, Greece

**Corresponding author: Giannis Zacharakis, email: zahari@iesl.forth.gr*

## Abstract

Advances in fluorescence microscopy have reached a plateau in improving depth-to-resolution ratios for imaging scattering tissues, highlighting an urgent need for innovative techniques. We introduce Wavelens, a groundbreaking opaque lens that combines dynamic wavefront shaping with a passive, angularly anisotropic photonic scattering plate—designed to scatter light predominantly in one direction. This compact, robust optoelectronic device, comparable in size to a conventional microscope lens, delivers superior light sheet formation capabilities. Using Wavelens, we achieved high-resolution LSFM imaging of diverse biological specimens, including 3D tumor spheroids and the intricate neural architecture of Caenorhabditis elegans in vivo. Notably, Wavelens maintains consistent light sheet properties even in dense, scattering-prone samples, overcoming a long-standing limitation of traditional microscopy. This innovative approach has the potential to revolutionize optical imaging, offering unprecedented clarity and resilience, even within dense, scattering-prone biological samples.

## Introduction

Precision light control is perhaps the corner stone for leveraging light to address global challenges and to create innovative solutions across diverse fields, including medical imaging, additive manufacturing, photolithography, cryptography and secure communications, fiber optics and adaptive lighting (e.g. automotive lighting systems). In medical and biological imaging, optical microscopy has been a foundational tool for centuries, despite being significantly constrained by light diffusion and the limited path control in tissue, which restricts its effective depth to superficial layers (<0.5mm). Due to photon diffusion, macroscopic optical imaging typically achieves only millimeter-scale resolution. This limitation is not exclusive to microscopy, but extends to all fields involving light interaction with optically opaque media. However, advancements in precise light path control within these media have the potential to fundamentally shift this paradigm, enhancing imaging depth and resolution across various applications.

A significant advancement toward precision light control is the concept of *wavefront shaping*[1–3] a transformative approach first theorized by Isaac Freund[4] and later experimentally implemented by Vellekoop and Mosk[5,6]. This innovative approach proposes that traditional linear optical elements can be replaced by disordered media. When combined with an active light control element, such as a spatial light modulator, disorder media enable any unitary transformation, marking a groundbreaking shift in optical manipulation capabilities[7]. Known as the *opaque lens concept,* this innovation holds broad implications across optics, with applications in adaptive optics, holography, ultrasound wave time-reversal, and telecommunications[3,8]. It has found its most profound application, however, in microscopy, where it helps overcome scattering challenges and reveals previously inaccessible details of biological structures. In microscopy, precision illumination can improve the depth to resolution ratio and significantly advance optical imaging capabilities beyond what is attained to date[9].

In microscopy, wavefront shaping addresses critical issues of penetration depth and resolution, enabling high-resolution imaging deeper within biological tissues and advancing our understanding of fundamental life mechanisms in vivo. A significant milestone in this area was the application of opaque lenses in microscopy providing the ability to reconstruct fluorescent objects concealed by turbid media, which illustrates the potential of this technology to reveal hidden biological details[10,11]

The opaque lens concept involves the integration of a spatial light modulator (SLM) with a scattering medium, enabling superior light modulation and focusing beyond conventional optics[12–14]. This is achieved by the capacity of the scattering medium to perturb the phase and amplitude of incoming waves, enriching the optical modes of the output field and enabling access to a wide range of optical degrees of freedom. The combination of the higher number of k-vectors with the SLM's light modulation capabilities allows precise light control, forming specific light structures at the opaque lens's output, marking a significant advancement in photonics.

Research into opaque lenses has demonstrated their capacity to focus light to a point smaller than the diffraction limit of conventional lenses[5,6] and to enhance the resolution of fluorescent microscopy beyond conventional methods[6,15]. Furthermore, reconstruction algorithms utilize the deterministic scattering of light and its inherent memory effect[16,17] enabling the reconstitution of objects obscured by scattering media[10,16].

However, to our knowledge, an opaque lens has not been functionally implemented in Light Sheet Microscopy, a powerful technique for high resolution imaging of large biological samples with minimal phototoxicity. Some preliminary studies[18,19] have shown that an opaque lens could be advantageous but the generated light sheets did not meet the minimum requirements for a realistic implementation in LSFM. As we are going to demonstrate, the bottleneck lies in the passive element of the opaque lens, and more specifically in the angular distribution of the scattered light. When a light beam propagates through a thin scattering plate, light is scattered in various directions. The angular distribution of the scattered light depends on the statistics and the strength of the perturbations of the scattering plate. These perturbations originate from microstructures that may involve thickness or refractive index variations (or both) that randomly occur in dimensions comparable to the wavelength. When there is no directional preference in these perturbations, light is scattered in an angularly isotropic distribution as shown in Figure 1(a). On the other hand, when there is a directional preference in these random perturbations, the scattered light distribution becomes angularly anisotropic. Such a case is depicted in Figure 1(b), where light is now scattered mainly along the horizontal $x$ direction.

Scattering media that exhibit highly angular anisotropy are key elements for the creation of light sheets using an opaque lens. By introducing a strong directional preference to the scattered light distribution, they conform to the lower dimensionality of a light sheet and enable efficient utilization of the complete angular spectrum components. This approach opens a new pathway to efficiently design novel light structures using opaque lenses by engineering the angular spectrum of a passive scatterer.

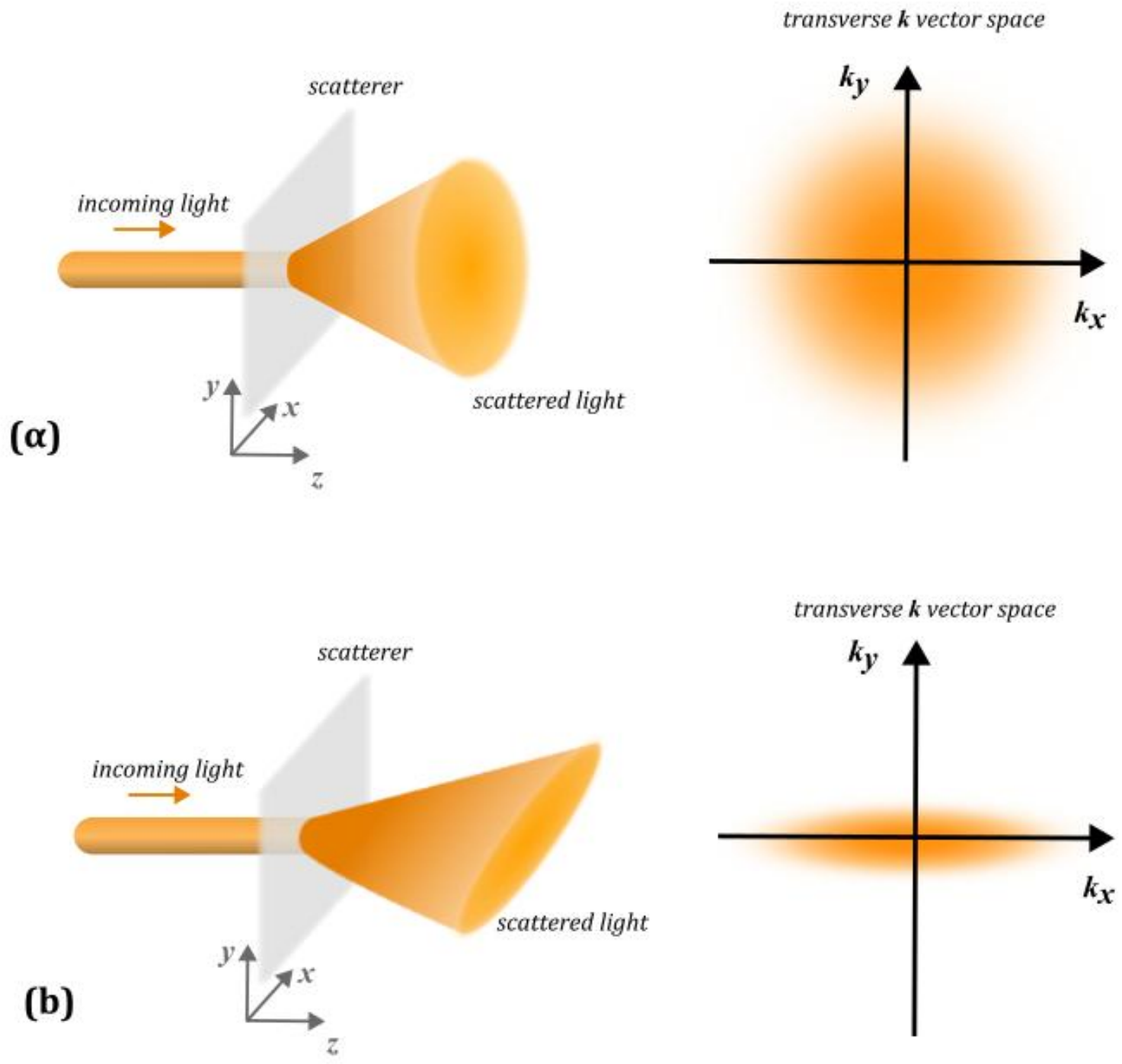

*Figure 1. Angular distribution of scattered light for various scatterers and corresponding angular spectrum. Left column: effect on a light beam, right column: power spectrum in transverse k-space of the scattered light distribution. (a) angularly isotropic scatterer, (b) angularly anisotropic scatterer*

Besides light sheet efficiency and quality, challenges remain in the widespread adoption of opaque lenses in microscopic imaging. These include among others, the ease of integration to existing modalities, robustness and repeatability of manufacturing and implementation. These factors have hindered the full potential of combining dynamic wavefront shaping with passive scatterers in practical applications[20,21]

To address these challenges, we introduce here Wavelens, a novel opaque lens that is designed as a compact, robust optoelectronic device, comparable in size to a traditional microscope objective lens. Optimized for generating high quality light sheets, Wavelens is equipped with a dynamic reflective mode SLM coupled with a passive, novel photonic scattering plate, GLAVKOS (*Glaucus Anisotropic k-Vector Optimization System*). This scattering plate is engineered to exhibit a strong directional preference in light distribution, enabling the precise control of scattered light for optimal light sheet formation. We demonstrate that, with these unique characteristics, the Wavelens can be optimized to generate and control high-quality light sheet structures suitable for applications in imaging large, live biological samples.

The capability of Wavelens as a high precision illumination device has been demonstrated in Light Sheet Fluorescence Microscopy (LSFM) for *in vivo* imaging of biological specimens, ranging from individual cells to entire model organisms. Using Wavelens, we have captured high-resolution images of live untreated tumor spheroids[22] and transgenic strains of *Caenorhabditis elegans*, showcasing its superior performance over conventional LSFM optics. Our results highlight Wavelens's potential providing crisper, less blurred images and underscoring its transformative impact on the field of optics and optical microscopy.

Wavelens represents a fully apochromatic, variable working distance excitation lens for microscopy, designed to provide high spatial resolution and virtually aberration-free light sheets. It addresses common stability issues, reduces sensitivity to environmental perturbations, simplifies alignment procedures and is easily tunable to meet application specific requirements. These features enable the use of advanced wavefront shaping methods, such as stepwise algorithms, offering a high precision solution for multiscale illumination applications.

## Results

### Wavelens optical performance

Figure 2 presents the optical performance analysis of the Wavelens optoelectronic device, specifically examining its capability to generate high-quality light sheet structures. Our experimental evaluation began by assesing the optical response consistency of the GLAVKOS medium, which is crucial for reliable light sheet-based imaging systems. To ensure this consistency, we fabricated five GLAVKOS scattering plates using a standardized method and tested each within the Wavelens device under identical experimental conditions. To compare the performance of Wavelens with light sheet formation through conventional optical elements, we conducted a series of tests using two standard cylindrical lens configurations, typically adopted in relevant microscopy systems. The first configuration involved a simple cylindrical lens with a +20 mm focal length, while the second configuration used a combination of an achromatic cylindrical lens (+50 mm focal length) and a 10X, NA = 0.28 plan apochromat microscope objective. A cross-sectional view of the produced light sheet structure through the simple cylindrical lens is presented in Figure 2a, revealing high-intensity side lobes surrounding the central light distribution. The observed pattern indicates the presence of a significant amount of spherical aberration in the tested optical configuration. In Figure 2b, the light sheet cross-section produced by the achromatic cylindrical lens and objective combination shows reduced side lobes due to partial aberration compensation, though these lobes still deviate from optimal performance. In contrast to the previous two configurations, the Wavelens-generated light sheet (Figure 2c) demonstrates a precise optical distribution with minimal aberration. For a direct comparison, we plotted the normalized intensity profiles of light sheets from *Figures 2a-c* against distance (\`). While all configurations produce light sheets of similar thickness (~1 μm), the Wavelens light sheet shows significantly reduced sidelobes, underscoring its superior quality and effectiveness in generating high-fidelity light sheets.

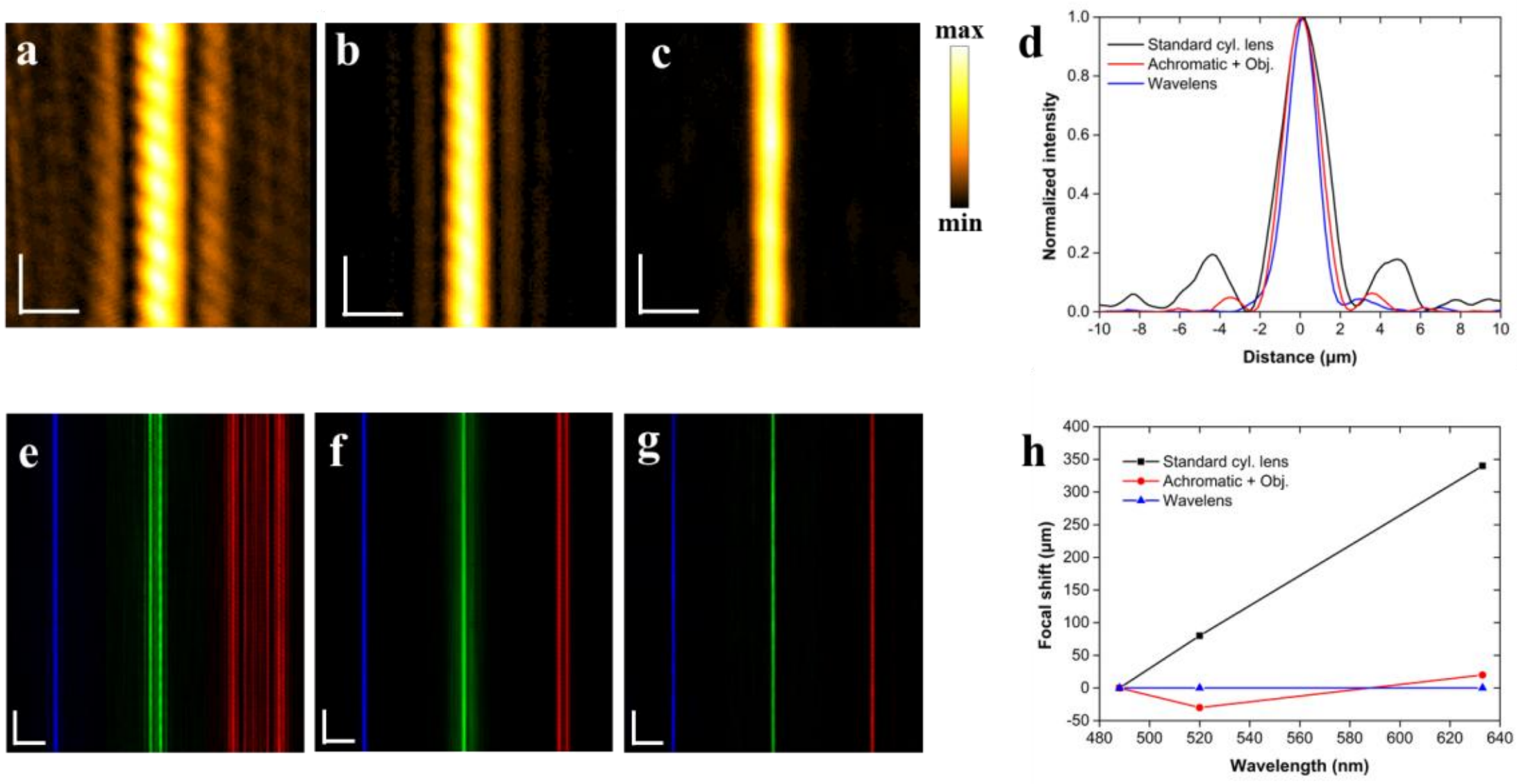

*Figure 2: Optical Performance Assessment and Multi-wavelength Evaluation of the Wavelens Device.*

*(a) Cross-sectional view of a light sheet produced by an uncorrected cylindrical lens. (b) Cross-sectional view of a light sheet produced by a combination of an achromatic cylindrical lens and a plan apochromat microscope objective. (c) Cross-sectional view of a Wavelens-generated light sheet, highlighting a well-confined, almost aberration-free spatial distribution. A 0,75 Gamma correction was applied in (a-c) for visualization purposes. Scale bar:5μm. (d) Normalized intensity profiles of the three light sheets presented in Figure 2a-c, emphasizing suppression of side lobes in the wavefront shaping approach. (e) Intensity cross-sections of light sheets formed using a fixed-position cylindrical lens at three different illumination wavelengths (488, 520, and 633 nm), indicating chromatic and spherical aberrations. (f) Cross-sections of light sheets formed using the combination of an achromatic cylindrical lens and a microscope objective at the same wavelengths, revealing reduced but still present chromatic and spherical aberrations. (g) Cross-sections of light sheets generated by the Wavelens device at all three wavelengths, showcasing identical focal planes and the absence of chromatic aberrations. In (e-g) scale bar: 20μm. (h) Graph plotting focal shift relative to the 488 nm focal plane due to chromatic aberrations for all three configurations.*

Figure 2e presents intensity cross-sections of light sheets formed using a fixed-position cylindrical lens at three different illumination wavelengths (488, 520, and 633 nm), highlighting the presence of both chromatic and spherical aberrations. The results in Figure 2f further illustrate that, despite using high-quality achromatic optics, a reduced but noticeable defocusing effect remains for red and green wavelengths. The residual chromatic aberration complicates the alignment and colocalization of multicolor light sheets. Additionally, spherical aberration is apparent in the characteristic striped pattern within the defocused beam intensity distribution. In contrast to previous setups with conventional optics, the Wavelens device effectively eliminates chromatic aberrations, focusing all three laser beams on practically the same focal plane (Figure 2g). To quantify these observations Figure 2h plots the chromatic aberration induced focal shift referencing the 488nm focal plane across all three configurations. For the first configuration (black line), the measured focal shift for 520 nm was observed at +80 μm, whereas the respective value for the 633 nm was found at +340 μm. The second configuration involving the achromatic lenses (red line), reduces these shifts to -30 and +20 μm for the same wavelengths, respectively. Notably, the Wavelens configuration (blue line) achieves a nearly identical focal plane for all laser lines, resulting in an almost negligible focal shift across the generated light sheets, found below the 5μm resolution limit of the system.

## Wavelens generated light sheets

In the next step, we evaluated the Wavelens's ability to generate ultrathin, sub-wavelength light sheet structures by utilizing a 520 nm laser line. Figure 3a shows a cross-section of a typical Wavelens-generated light sheet recorded at the plane of its formation through the CCD camera of the system with a field of view measuring 135 by 110 μm$^2$. The resulting image depicts a well-defined light intensity distribution with high contrast against the background (SBR =100), confirming the optimized performance of Wavelens under such conditions. The normalized intensity profile of the light sheet shown in Figure 3b (black data points) has been further analyzed by a double Gaussian model fit (red curve), yielding a full-width at half-maximum (FWHM) value of 478 nm ($R^2$ = 0.92), which is clearly narrower than the 520 nm illumination wavelength.

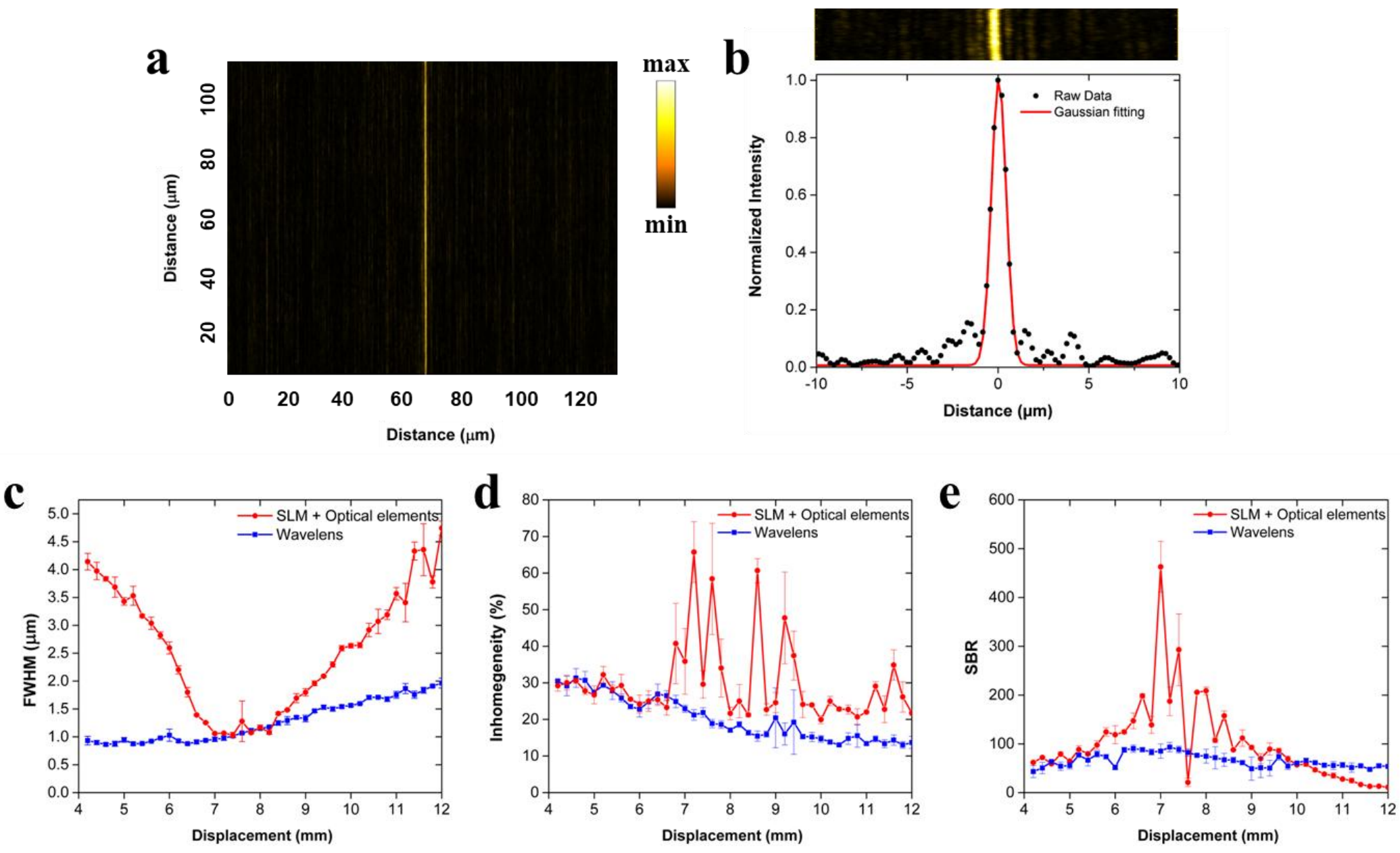

*Figure 3. Comparative analysis of light sheet generation and control through Wavelens and SLM-assisted conventional optics.*

*(a) Cross-sectional view of a Wavelens-generated light sheet, captured by the CCD camera at the formation plane, showcasing a high-contrast intensity distribution against the background. Field of view: 135 x 110 μm². (b) Normalized intensity profile of the Wavelens light sheet (black data points) fitted with a double Gaussian model (red curve), yielding a Full Width at Half Maximum (FWHM) of 478 nm, significantly below the 520 nm illumination wavelength. (c) Graph comparing average light sheet FWHM values as a function of focal displacement for both the SLM-assisted conventional optics (red curve) and the Wavelens device (blue curve), highlighting the extended efficiency of Wavelens over a*

*working distance range of 8 mm. (d) Graph illustrating the inhomogeneity of light sheet profiles at variable working distances for both configurations, showcasing Wavelens superior uniformity and control. (e) Graph of the Signal-to-Background Ratio (SBR) of the light sheet profiles as a function of working distance, demonstrating the consistent performance of Wavelens across the focal displacement range, as opposed to the variable SBR in SLM-assisted configurations.*

To further demonstrate the superior capabilities of Wavelens in effective light control and manipulation, we compared it with an optical configuration involving the combination of a SLM and conventional optics. With this comparison we particularly investigated and demonstrated that the integration of GLAVKOS anisotropic scattering medium is crucial in creating stable, high quality light sheets across extensive working distances. Initially we used an SLM with an achromatic lens setup to generate light sheets spaced 15 μm apart in the XY plane, shifting the working distance slightly from the focal point. We followed a similar process with Wavelens, extending the working distance up to 8 mm to assess its range.

Figure 3c compares the average light sheet FWHM values as a function of focal displacement for both the SLM-assisted conventional optics configuration (red curve) and Wavelens device (blue curve). For each position the average FWHM values have been calculated from the three generated light sheets, with error bars corresponding to the standard deviation of the respective measurements. While the SLM-assisted configuration achieves relatively thin light sheets (between 1 and 2 μm) for a limited working distance range not exceeding 3 mm. However, in the case of displacements beyond this, we observe a steep increase of the measured FWHM value as a function on the working distance, revealing a limited light control efficiency in these regions. Wavelens, by contrast, maintained thin and stable light sheets across a 2.6 times larger range approximating 8 mm, with smoother performance attributed to GLAVKOS's optimized scattering properties.

Moreover, Figure 3d illustrates the inhomogeneity of the light sheet profile at variable working distances for both configurations. The apparent strong fluctuations and the large error bars associated with the SLM-assisted configuration indicate the limitations in effectively controlling the light, particularly near the original focal plane. In contrast, Wavelens exhibits consistently a high degree of light sheet uniformity along the evaluated displacement range, emphasizing the effectiveness in light manipulation due to the integration of the GLAVKOS medium.

Finally, Figure 3e depicts a plot of the signal to background ratio (SBR) of the light sheet profile as a function of working distance. Although the SLM-assisted configuration produces a higher SBR light sheet structure near the focal plane, albeit with reduced stability beyond that, the Wavelens device maintains a relatively constant SBR across the entire focal displacement range. These results align with previous Wavelens metrics, supporting its superior light manipulation capabilities and performance for Light Sheet Fluorescence Microscopy (LSFM) over configurations combining SLM and conventional optics without the integration of a specialized scattering medium.

**Proof of concept imaging of live fluorescence samples**

Having characterized the properties of the light sheet structures generated by Wavelens, we continued in evaluating its imaging performance when imaging biological specimens. Specifically, we employed a three-dimensional tumor spheroid composed of U87MG glioblastoma cells expressing Green Fluorescence Protein (GFP) in their cytoplasm and nucleus. Using our in-house build light sheet fluorescence microscope, we compared the Wavelens's output to that of a conventional optical setup with a cylindrical lens and a 10X objective, which served as a reference. For the comparison we utilized a 488 nm excitation line to image the spheroid on the second day post-formation and then visualize a cross-sectional view at a depth of 85μm. The reference image of the spheroid at the XY and YZ cross sectional planes are shown in Figure 4a and 4b, respectively, depicting several bright spots corresponding to individual cells. Nevertheless, a diffuse fluorescence background permeates the entire spheroid region, diminishing image clarity and contrast. In contract, imaging data acquired using the Wavelens , are presented in Figure 4c and 4d, respectively, where we clearly observe a significant enhancement in the performance both in terms of contrast, spatial resolution and sensitivity as regards the detection of single cells. The background fluorescence was substantially reduced, yielding clearer cellular outlines and enhancing the distinction between cellular structures and surrounding areas.

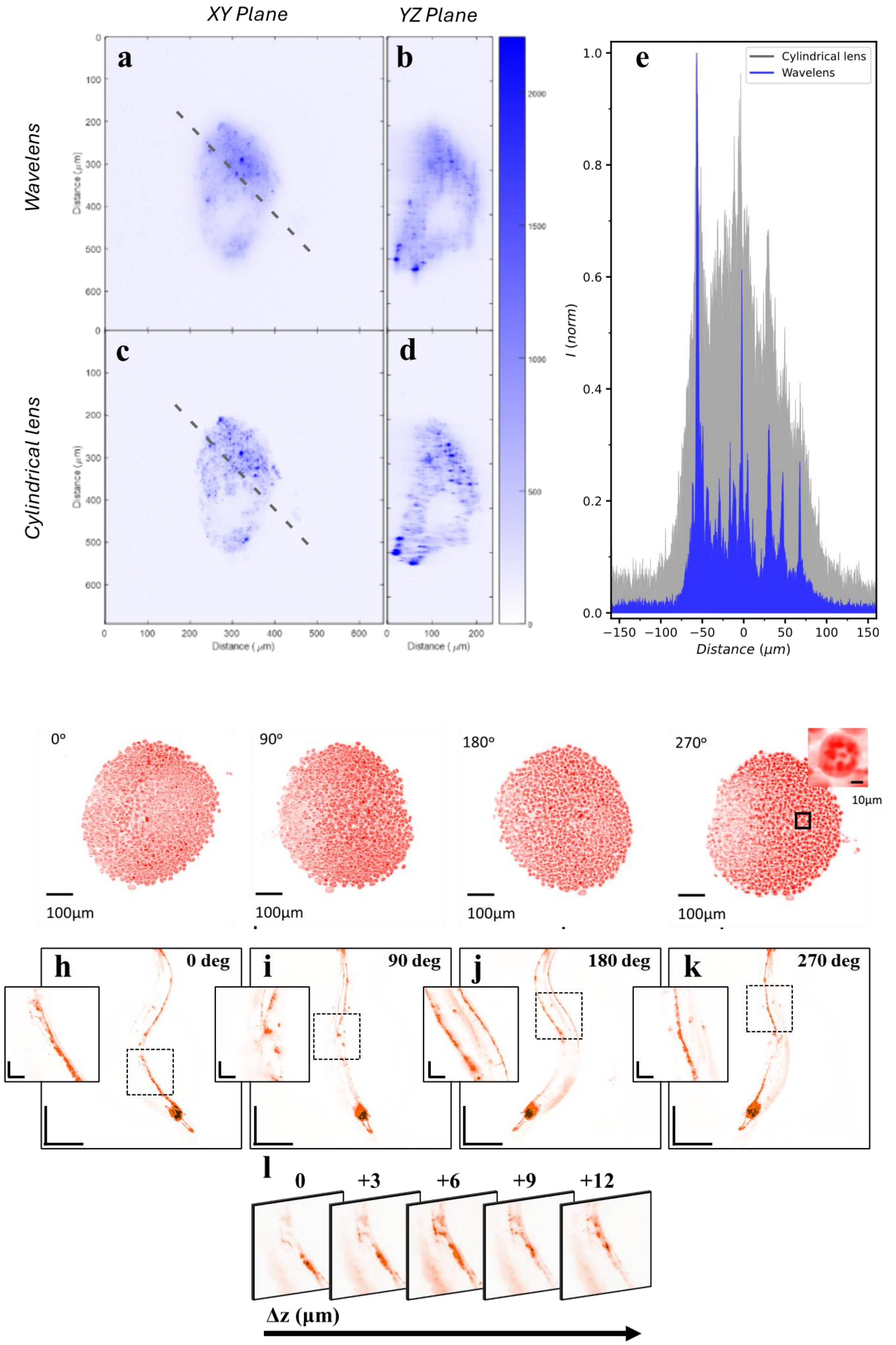

*Figure 4. Evaluation of Wavelens-based Light Sheet Fluorescence Microscopy (LSFM) for in vitro and in vivo imaging. Subfigures 4a-b show cross sections of a GFP-expressing U87MG glioblastoma*

*spheroid with conventional optics at 85μm depth and subfigures 4c-d the corresponding images acquired with Wavelens, highlighting its improved performance in terms of spatial resolution, and single-cell detection capabilities. Figure 4e displays the normalized pixel intensity profiles along the dotted lines in 4a and 4c , revealing the enhanced feature distinguishing capability of Wavelens. Subfigure 4f-g present four-angle maximum intensity Z-projections of a DRAQ5-labelled spheroid, showcasing the high-resolution visualization of glioblastoma cell nuclei. Scale bar:50μm. Figures 3h-k illustrate LSFM imaging of a DsRed2::LGG-1-expressing Caenorhabditis Elegans (Scale bar: 150μm), providing an overview of the nervous system and highlighting neuronal spatial details within a selected 172 by 172 μm² ROI. Scale bar: 30um. Finally, Figure 3l presents a series of sequential cross-sectional images of the nematode, highlighting the optical sectioning capability of the Wavelens device.*

To further demonstrate the improved imaging quality offered by Wavelens, we have generated normalized pixel intensity profiles across the indicated dotted lines from the data shown in Figures 4a and 4c. The profiles are depicted in Figure 4e for a distance of 300 μm, revealing the capability of the Wavelens to resolve fine details otherwise hidden bellow the background fluorescence signal in the reference image (grey curve; one peak detected). Moreover, the full capabilities of Wavelens-based LSFM imaging *in vitro* have been assessed using a similar tumor spheroid with DRAQ5 nuclear labelling at the 7th day post-formation. Figures 4d-g present the maximum intensity Z-projections of the spheroid acquired at four orthogonal angles (0, 90, 180 and 270 degrees respectively), allowing for the visualization of the glioblastoma cells' nuclei with sufficiently high resolution in different views.

In our final imaging assessment, we used Wavelens for *in vivo* imaging of transgenic *Caenorhabditis elegans*, expressing the fluorescent marker DsRed2::LGG-1 in the neurons. In particular, DsRed2::LGG-1 serves as a marker for autophagosomes and intracellular vesicles (300-500 nm in size), which are crucial for cellular degradation via autophagy. By utilizing Wavelens-enabled light sheets, we aimed to capture high-resolution, extended field-of-view images of the nematode's nervous system. Maximum intensity Z-projections of an anesthetized *C. elegans* nematode were obtained from four orthogonal angles as presented in Figures 4h-k. The reconstructions provide detailed images across the animal's entire nervous system, whereas the respective insets at the left part of the figures highlight spatial details of the neurons at the sub-cellular level, with individual dots representing autophagosomes, within a selected ROI of 172 by 172 μm² indicated with the dotted squares. Finally, Figure 4l presents a series of five sequential cross-sectional (slice) images of the animal at 0 degrees, captured along the scanning direction with a step size of 3 μm. The high-quality light sheet provided by Wavelens (Figure 3a) enables us to clearly resolve the individual characteristics of the neurons.

## Discussion

Optical microscopy has been foundational in accelerating biological imaging by providing a powerful tool for visualizing life and disease. Yet despite recent breakthroughs in advancing its capabilities, its development is still constrained by the achievable depth-to-resolution ratio within scattering tissues[9]. Light diffusion inside biological tissue limits the ability to capture high-resolution images to a few mean free paths, restricting optical microscopy to superficial (<0.5mm) investigations. Recent innovations have created a promising scientific frontier in overcoming this obstacle, by the integration of wavefront shaping techniques, offering unprecedented control over light propagation in highly scattering environments effectively countering multiple scattering events to enable deeper, more precise imaging within complex biological tissues[1,2].

Nevertheless, despite the transformative impact of wavefront shaping, its practical applications are limited by the available spatial modulation efficiencies of current optical configurations[13]. A solution for overcoming these boundaries in light manipulation, is provided by the synergy between engineered disordered photonic media and sophisticated wavefront shaping strategies[12,23] These configurations are also referred to as opaque lenses[24]. Expanding on this concept, we introduce Wavelens, a complete wavefront shaping device that incorporates GLAVKOS, a passive, angularly anisotropic scattering medium. Designed and implemented to be as compact as a traditional microscope objective lens, Wavelens can be seamlessly integrated into any Light Sheet Fluorescence Microscopy (LSFM) system and digitally control illumination light and streamlining alignment processes, while diminishing the adverse effects of environmental variability on imaging quality.

Previous attempts to generate light sheets through wavefront shaping, in conjunction with uniform disordered media, have been hindered by fabrication complexity and light manipulation efficiency[13,23,25]. Our new scattering medium, GLAVKOS, however, breaks through the above limitations by achieving enhanced angular anisotropy in scattering, increasing the available angular spectrum components, which is essential for efficient light control. Exploiting these unique properties, Wavelens is able to generate light sheets with superior quality and precision, characterized by sub-wavelength full width at half maximum (FWHM), high intensity homogeneity and improved signal to background ratio (SBR). These attributes are critical for enhancing spatial resolution, reducing artifacts, and improving the signal to noise ratio (SNR) in obtained images. Notably, the characteristics of GLAVKOS allow for exceeding previous benchmarks set by anisotropic photonics glasses APG[18], achieving a minimal light sheet width of 500 nm and a signal to background ratio exceeding 100. Furthermore, GLAVKOS exhibits a transmission efficiency above 80% when compared to free space propagation, effectively addressing the common power losses associated with wavefront shaping techniques.

Another significant improvement of light sheet formation through Wavelens is the mitigation of spherical and chromatic aberrations, a common drawback in conventional optical setups. This capability is crucial for multicolor LSFM systems, where precise, co-localized focusing is essential. Wavelens is also very effective in generating ultra-thin light sheets, facilitating the detection and visualization of minute spatial details and offering unmatched control over light sheet formation across extended working distances. The latter feature is indispensable for examining large biological samples without compromising spatial resolution or image quality.

Employing our in-house developed Wavelens-enabled light sheet microscope we have managed to image live biological samples both *in vitro* and *in vivo*, demonstrating the advancement in imaging quality offered by our approach. We have selected to study two models, each with specific significance in terms of both biological relevance and challenging imaging requirements. The first model consists of cancer spheroids with high capacity to mimic the intricate complexities of *in vivo* tumors, yet they present significant challenges when imaged due to their optical opaqueness. Our study has demonstrated significant advancement in imaging capabilities with the use of Wavelens, offering increased resolution throughout the entire volume of the spheroids. When compared and benchmarked to the performance of conventional optical configuration the Wavelens-enabled LSFM presents a clear advantage in terms of resolving subcellular spatial details with minimal background fluorescence signal.

Our second model, a transgenic *C. elegans* nematode, presents challenges in obtaining high resolution images at large fields of view when imaged with confocal microscopy. Our system has managed to image the entire nematode at subcellular resolutions resolving specific fluorescence emission from neuronal autophagosomes, which are intracellular components with sizes between 300nm - 500nm. *C. elegans* expressing the fluorescent marker DsRed2::LGG-1 in the neuronal autophagosomes serves as an important model for studying the cellular processes involved during ageing, signifying the role our approach can play in better understanding ageing and its effect at the organism level.

With its unique capabilities, Wavelens represents a significant breakthrough in light sheet fluorescence microscopy and potentially in the wider microscopy field, offering high-contrast, high-resolution imaging deeper in complex biological samples. Its advanced light manipulation capabilities allow Wavelens to deliver consistently high-quality light sheets over extended working distances, reducing background noise, enhancing sensitivity and making it particularly effective for visualizations and analysis at the cellular and subcellular level. Finally, Wavelens holds significant potential for advancing biological imaging, offering an essential tool for applications that demand precision light control for high fidelity imaging.

## Materials and methods

### Wavelens-based imaging system

The Wavelens-based imaging system, purposefully named WaveSPIM (Fig. 5a), employs three continuous wave (CW) laser beams emitting at 488, 520 and 633 nm, respectively, to achieve efficient specimen illumination through the generation of optimized light sheet structures. A linearly polarized beam is initially expanded and subsequently directed towards the Wavelens device, which shapes the wavefront by modulating the phase of the incident optical radiation to attain light sheet illumination at user-defined positions. To this end, Wavelens (Fig. 5b) combines a reflecting phase SLM device, a beam splitter, two prisms, a cylindrical lens and finally GLAVKOS, which constitutes a highly angular anisotropic scattering medium. Following its transmission through Wavelens, the resulting elongated speckle pattern is collected by an objective lens and imaged on a CCD camera. A wavefront optimization process controlled by a stepwise iterative algorithm (Supplementary Fig. 2) adjusts the Wavelens response using optical feedback from the CCD. Specifically, an initial phase mask is generated and projected on the Wavelens SLM device, in order to shape the wavefront. The resulting scattered light pattern is captured by the CCD and is subsequently compared with a pre-calculated target image, representing the desired light sheet structure, through the estimation of the Pearson correlation coefficient[26]. The phase mask is iteratively modified at pre-defined discrete phase levels until the maximum correlation with the target image has been achieved. The resulting light sheet is then utilized for efficient illumination of biological samples in LSFM measurements. Prior the imaging session, the biological samples are embedded in an agarose solution and placed inside a custom-designed bath. Precise positioning of the samples is achieved using a motorized motion control system capable of four-axis translation and rotation. The recording of the fluorescence images is performed through a second objective lens placed perpendicular to the illumination arm, an optical filter to cut-off the excitation light, and a scientific CMOS camera for the sensitive detection and recording of the signals.

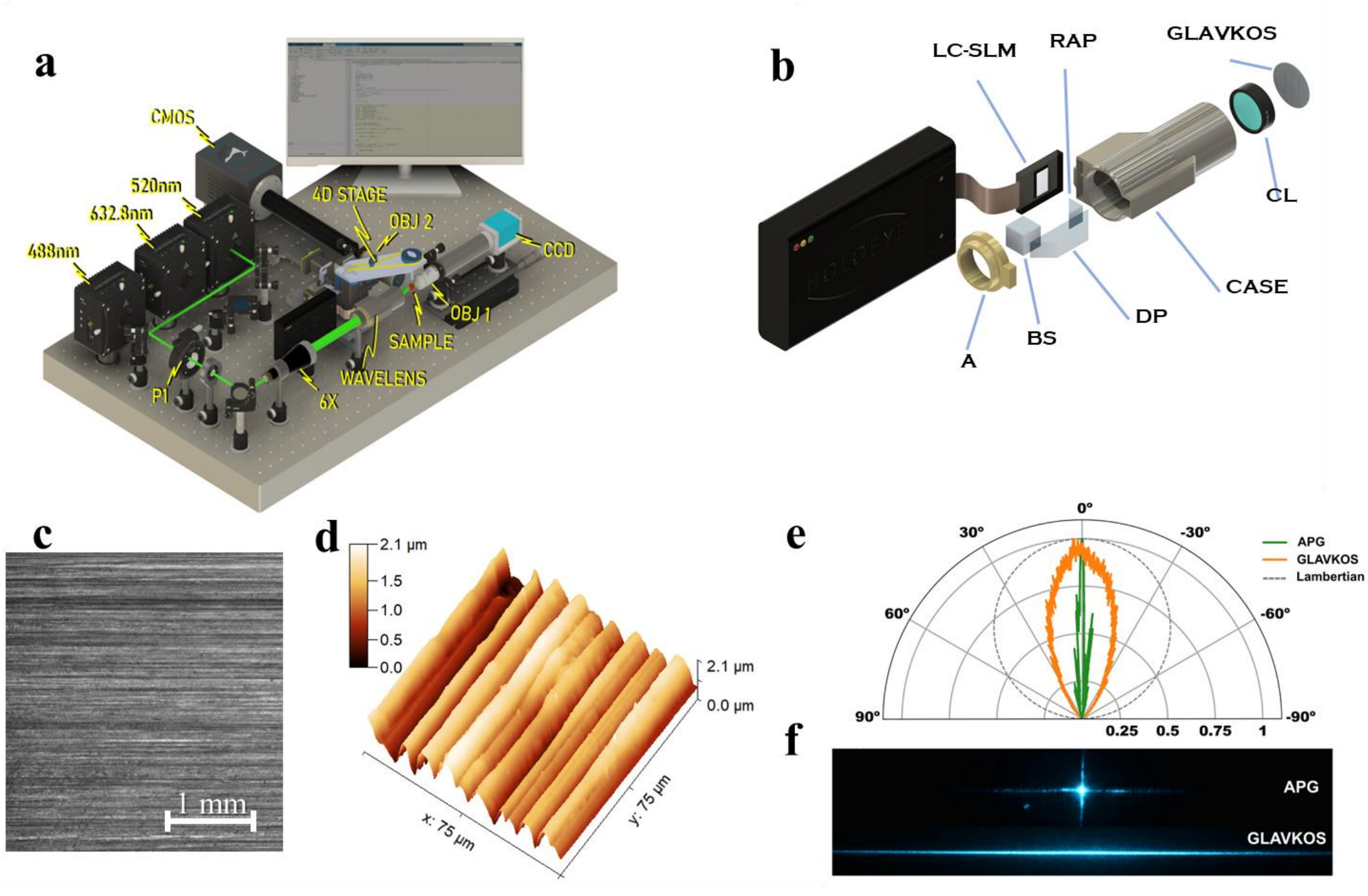

*Figure 5: Schematic of the Wavelens-based Imaging System and Characterization of GLAVKOS Scattering Medium. (α) Graphical representation of the optical setup utilized for Wavelens*

*characterization and Light Sheet Fluorescence Microscopy (LSFM). Abbreviations: M1, Mirror; FM(1-2), Flip Mirrors; P1, Linear polarizer; BE, Beam expander; BE, Beam expander; OBJ(1-2), microscope objectives; CCD, Optimization camera; Z stage, Z-axis translation stage; FF, Fluorescence filter; CMOS, Fluorescence detection camera. (b) Wavelens graphical representation. Abbreviations: A, Threaded adapter; BS, Beam splitter; LC-SLM, Liquid crystal spatial light modulator; DTP, Dove tail prism; RAP, Right angle prism; CL, +20mm Cylindrical lens; GLAVKOS, Glavkos scattering medium; CASE, optoelectronic parts casing. (c) High-resolution brightfield image showing the dense groove structures in GLAVKOS. (e) AFM image of GLAVKOS at a characteristic area of 75x75μm (e) Polar plot of light distribution after transmission through the APG medium, GLAVKOS compared to a Lambertian distribution. (f) Comparative image of light distribution after transmission through the APG medium, GLAVKOS .*

## Characterization of GLAVKOS scattering plate

Structural characterization of the GLAVKOS scattering medium has been accomplished by optical and atomic force microscopy (AFM) imaging. A high-resolution brightfield image of GLAVKOS (Fig. 5c) reveals the formation of spatially dense, highly directional groove structures in the form of parallel horizontal stripes, which have been produced through the mechanical etching of a polymeric substrate. The parallel grooves of GLAVKOS diffract the illuminating light in a random manner according to their individual spatial features. As a result, the stripes in Fig. 5c and Fig, 5d appear in an arbitrary distribution emerging from the inherent randomness of the scattering medium in terms of the grooves' spatial density, width and depth. Complimentary to the optical, an AFM image is presented in Fig. 5d recorded at an area *of 75x75μm*, explicitly illustrating the smooth elongation and the uni-directionality of the grooves. The above observations render GLAVKOS a highly angular anisotropic scattering medium. Optical and AFM images demonstrate that in any scale the groove distribution is random with high variability in the geometrical features of the etched structures. The average spatial period of the grooves, has been measured at 5 μm.

Aiming to provide a direct comparison of light scattering behavior between GLAVKOS and a state-of-the-art APG medium, which was previously utilized[18] for wavefront shaping, we have employed a TEM00 HeNe laser (Newport, N-LYP-173) emitting at 594 nm for the effective illumination of both structures. Fig. 6e demonstrates the angular distribution of the scattered light intensity along the horizonal direction of the APG, GLAVKOS and a typical Lambertian. Furthermore, the resulting light pattern on a flat screen located 20 cm after the scattering plates is shown in Fig. 6f. GLAVKOS presents a significantly enhanced scattering efficiency with strong angular anisotropy. Light is scattered in the horizontal direction in a cone angle that is 62x larger compared to that in the vertical direction. On the other hand, APG scatters light to a much narrower angular range, with no directional preference. Side lobes that are observed and the cross-like structure in the image are due to light diffracted from the edges of the APG structure.

**Experimental setup description**

The experimental setup (Figure 5a) employs three continuous wave (CW) laser beams at 488 nm (Coherent, Sapphire 488-200 CW CDRH, 200mW), 520 nm (Thorlabs, L520P50, 50mW), and ~633nm (Melles Griot, model 05-LHR-991, 30mW). The desired laser beam is selected using a combination of a mirror (M1) and two flip-mount mirrors (FM1 and FM2). The polarization axis of the beam is defined by a polarizer (P1; Thorlabs-LPVISE100-A) as the outcoming laser light is not completely linearly polarized. To expand the laser beam, a beam expander (BE) has been utilized, which magnifies the laser beam by a factor of 6x. The expanded laser beam is then directed towards the Wavelens, which in turn shapes the wavefront by modulating the phase of the incident light. The scattered light after the Wavelens is collected using a microscope objective lens (OBJ1; Edmund DIN 20X/0.4 NA) and imaged on a CCD camera (CCD; Thorlabs, DCU224M). The wavefront optimization process is performed by an iterative step wise algorithm running on a PC (see Iterative algorithm section), which controls the Wavelens and utilizes optical feedback from the CCD for optimization purposes. Once an optimal light sheet structure is formed, it can be exploited for efficient specimen illumination in LSFM measurements. The sample is embedded in 0.8 % w/v agarose solution before placed inside a custom-designed 3d-printed bath filled with distilled water. For precise four-axis sample positioning (XYZ translation and rotation), we employ a commercially available motorized motion control system (Picard, USB 4D Stage). Finally, a second microscope objective (OBJ2; Mitutoyo 20X Plan Apo Infinity Corrected Long WD) in combination with an appropriate fluorescence filter FF and a scientific CMOS camera (Hamamatsu, ORCA-Fusion C14440-20UP) are used for fluorescence signal detection.

**Wavelens description**

The Wavelens optical elements and their assembly order are shown in Figure 5b. Initially, the incoming beam is incident on a beam splitter (BS; Thorlabs BS004) and split into two parts of equal intensity. While the radiation transmitted through the BS is blocked, the reflected portion is directed towards a SLM screen (LC-SLM; Holoeye, PLUTO-2.1 LCOS), which reflects back the phase-modulated light through the BS into a dove prism (DP; Thorlabs PS991). Following two successive reflections within the DP, the modulated radiation is guided into a right-angle prism (RTP; Thorlabs PS910), restoring the initial direction of the beam before the BS incidence (linear optical path design). Subsequently, in order to achieve a higher coupling efficiency with the SLM mask, an $f = +20$ mm cylindrical lens (CYL; Thorlabs LJ1960L1-A) is used to focus the light on the front surface of the anisotropic scattering medium (GLAVKOS). A threaded adapter is used to integrate the Wavelens device on standard optical setups. The optical and optoelectronic components are enclosed in a 3d-printed housing (CASE), fabricated in-house using PLA as building material. The total length of the device is 83 mm, and its maximum width is approximately 41 mm. Furthermore, the mount of the body offers the capability to fine-adjust the direction of the light beam, achieving the maximum imaging efficiency.

## Fabrication of the GLAVKOS scattering medium

Glaucus Anisotropic k-Vector Optimization System (GLAVKOS), is a novel passive optical element that has been fabricated using a succession of mechanical and chemical processes resulting in a unidirectional random surface etching. The surface relief achieved through this process is of sufficient density and magnitude to lead to angularly anisotropic light scattering, at a preferential direction and over an extended k-vector range. The material used to construct this element provides a sufficiently high refractive index contrast value and high transparency in the optical frequency domain.

GLAVKOS is fabricated using the following procedure: First, a suitable optically transparent plate, for example PMMA with a thickness of 0.6mm is selected. The plate surface is then carefully cleaned using isopropyl alcohol or a suitable cleaning solvent to ensure the removal of any dust or debris. The next step involves choosing the appropriate abrasive that will mechanically etch the surface and result in the desired surface roughness. The plate is mechanically secured preventing any movement during the mechanical etching process. The process begins by applying constant pressure with the abrasive while moving it along the desired direction. This procedure creates uniaxial grooves on the surface, which are crucial for achieving angular anisotropic scattering. Depending on the desired effect, multiple passes with the abrasive may be necessary to achieve the desired depth and statistical uniformity of the grooves. Once the etching is complete, the surface is cleaned again to remove any debris generated during the process. The scattering plate is then tested in respect to its scattering strength and angular anisotropy and the mechanical process is repeated if necessary.

**Iterative algorithm**

The recorded signal is fed into the iterative algorithm, which modulates the beam to achieve the desired output. The resulting light structure is captured by a CCD camera, which is used for optical feedback. As a preliminary step a target image that represents the desired outcome is defined. The step wise optimization process for focusing commences by setting the phase of the spatial light modulator (SLM) to zero. Due to the time-consuming nature of exploring all degrees of freedom of the SLM, a subset of relevant parameters (e.g., height, width, phase levels) is chosen for modulation. The algorithm then calculates the Pearson correlation[26] between the current state image and the target image, providing a quantitative measure of their dissimilarity. An increase in correlation values indicates convergence towards the target image. The SLM mask is then configured by dividing it in $N$ super-pixels, with a shape that relates to the device modality. Each super-pixel is set to the same phase value, which ideally spans from 0 to $2\pi$, with discrete phase values determined by the bit depth of the phase SLM device. Following the target setting and correlation calculation, a routine is initiated, wherein the algorithm tests $M$ different phase shift values interpedently for super-pixel and computes a new correlation value, while the rest super-pixels are set to zero phase. The algorithm keeps a phase value for a super-pixel only if the measured correlation is higher than the currently stored correlation. Otherwise, the previous configuration is restored, and the algorithm proceeds to the next super-pixel. This routine is repeated for each of the $N$ super-pixels.

**Preparation of biological samples**

*C. Elegans*

The *C. elegans* strain that we used in our study was the IR2379 (N2; Ex[rab-3p::dsRed::lgg-1]) expressing DsRed in neurons. This strain was created in the NT lab and first described in this study. We used standard methods for *C. elegans* maintenance. Nematodes were cultured at 20ºC, on NGM streptomycin-supplemented plates, seeded with OP50 Escherichia coli bacteria as a food resource. Young adult worms can be immobilized in 10 mM levamisole (Sigma-Aldrich, St. Louis, USA)/M9 buffer and mounted in 2% agarose solution. Levamisole temporarily suppresses muscular activity by blocking the function of nicotinic acetylcholine receptors (nAChR). In the applied concentration and incubation time this drug is not lethal for *C. elegans*. Animals were mounted into a capillary (QuikRead go® 10 µl, Aidian, Eschborn, Germany) in 0.8% low-melting agarose (Thermo Fisher Scientific, USA). The capillary was carefully cleaned with 70% ethanol.

*Tumor spheroids*

The home-established patient-derived Glioblastoma (GBM) cell line GBP03[27] was used in this study. The cells were cultured in Dulbecco's modified Eagle medium (DMEM; Gibco, UK) supplemented with 10% heat-inactivated fetal bovine serum (FBS) and 50µg/ml gentamicin (PANREAC Applichem, Germany; a.k.a. DMEM++) and incubated in standard lab conditions (37 °C, 5% CO2, 95% humidity).

For the generation of 3D multicellular spheroids, approximately 200 single cells per 100µl DMEM++ were plated in an ultra-low binding U-bottom 96 well plate (Nunclon Sphera, Thermo Scientific). LSFM imaging was performed on the day of spheroid formation (designated as Day 2). The cells were genetically labeled with the green fluorescent protein (GFP) and the cell nuclei were counterstained with the fluorescent probe DRAQ5 (Abcam, UK). Probe labeling was performed approximately 2 hours before imaging. The spheroids were mounted in capillary tubes (QuikRead go® 10 µl, Aidian, Eschborn, Germany) in 0.8% low-melting agarose (Thermo Fisher Scientific, USA).

## Acknowledgements

The Authors would like to acknowledge the financial support by the Hellenic Foundation for Research and Innovation (HFRI) under the HFRI PhD Fellowship grant (Fellowship Number 488), the H20202 FETOPEN project "Dynamic" (EC-GA-863203), the FET Innovation Launchpad project "Wavelens" (EC-GA-101035014), the EIC Pathfinder Open project "SWOPT" (EC-GA- 101046667) and the NSRF 2014-2020 "BIOIMAGING-GR" (MIS5002755) project.

## Author contributions

Conceptualization: Mikis Mylonakis, Evangelos Marakis, Athanasios Zacharopoulos, Giannis Zacharakis ; Methodology: Mikis Mylonakis, Evangelos Marakis; Software: Evangelos Marakis, Mikis Mylonakis; Validation: Mikis Mylonakis, Maira Tampakaki, Angela Pasparaki, Nektarios Tavernarakis; Formal analysis: Evangelos Marakis, Mikis Mylonakis, Iro Manne; Investigation: Mikis Mylonakis, Evangelos Marakis; Resources: Maira Tampakaki, Angela Pasparaki, Nektarios Tavernarakis, Giannis Zacharakis; Data Curation: Evangelos Marakis, Mikis Mylonakis; Writing - Original Draft: Mikis Mylonakis, Evangelos Marakis, George J. Tserevelakis; Writing - Review & Editing: George J. Tserevelakis, Giannis Zacharakis, Nektarios Tavernarakis, Dimitrios G. Papazoglou; Visualization: Mikis Mylonakis, George J. Tserevelakis, Evangelos Marakis; Supervision: George J. Tserevelakis, Giannis Zacharakis, Dimitrios G. Papazoglou; Project administration; Giannis Zacharakis; Funding acquisition: Giannis Zacharakis.

## Competing interests

The author(s) declare no competing interests.

## Data availability

The datasets generated and analyzed during the current study are available from the corresponding author on reasonable request.

## Supplementary Notes

### Experimental setup

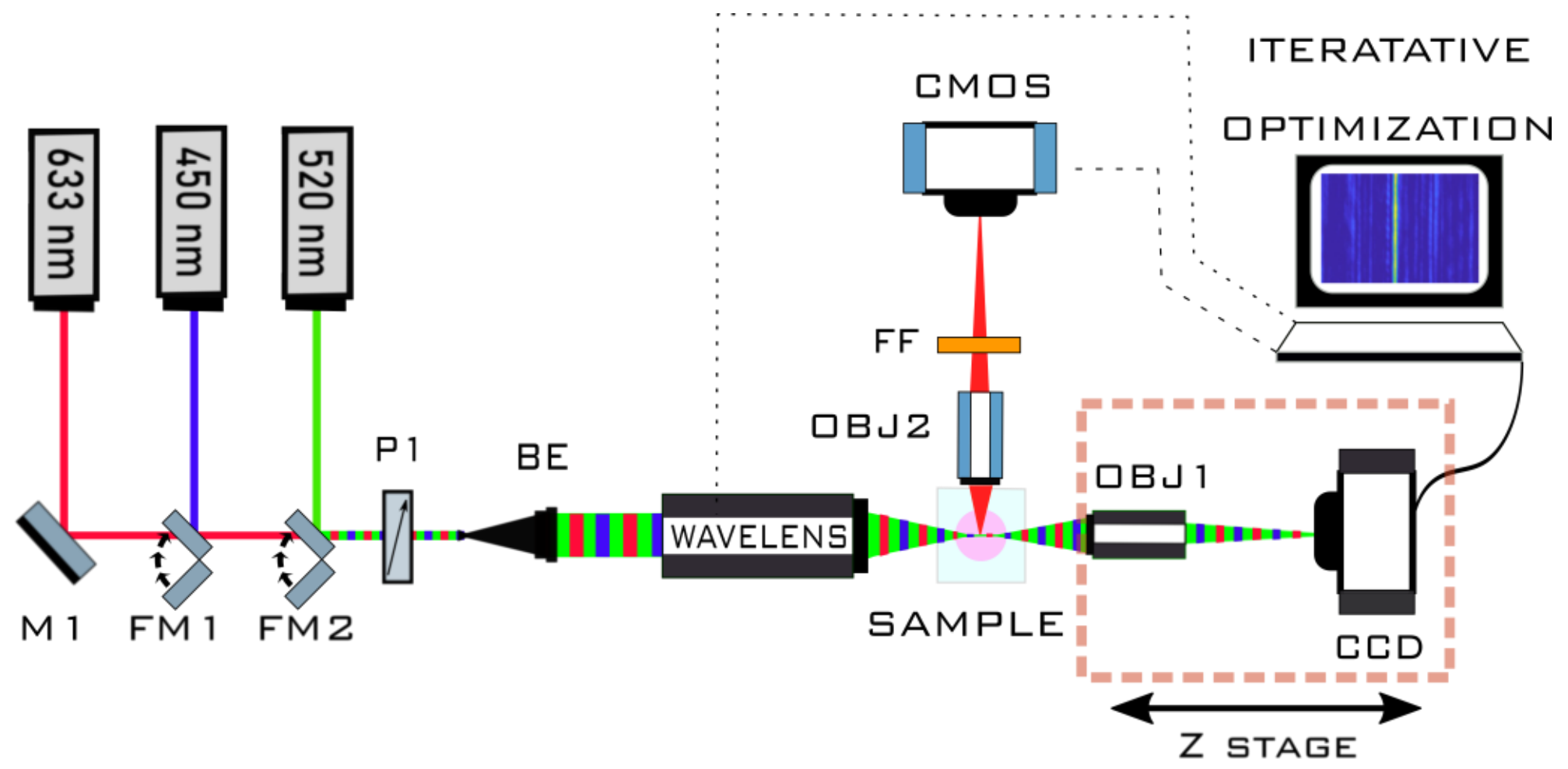

*Supplementary Figure 1. Schematic of the Wavelens-based Imaging System and Characterization of GLAVKOS Scattering Medium. (α) Graphical representation of the optical setup utilized for Wavelens characterization and Light Sheet Fluorescence Microscopy (LSFM). Abbreviations: M1, Mirror; FM(1-2), Flip Mirrors; P1, Linear polarizer; BE, Beam expander; BE, Beam expander; OBJ(1-2), microscope objectives; CCD, Optimization camera; Z stage, Z-axis translation stage; FF, Fluorescence filter; CMOS, Fluorescence detection camera.*

### Iterative algorithm

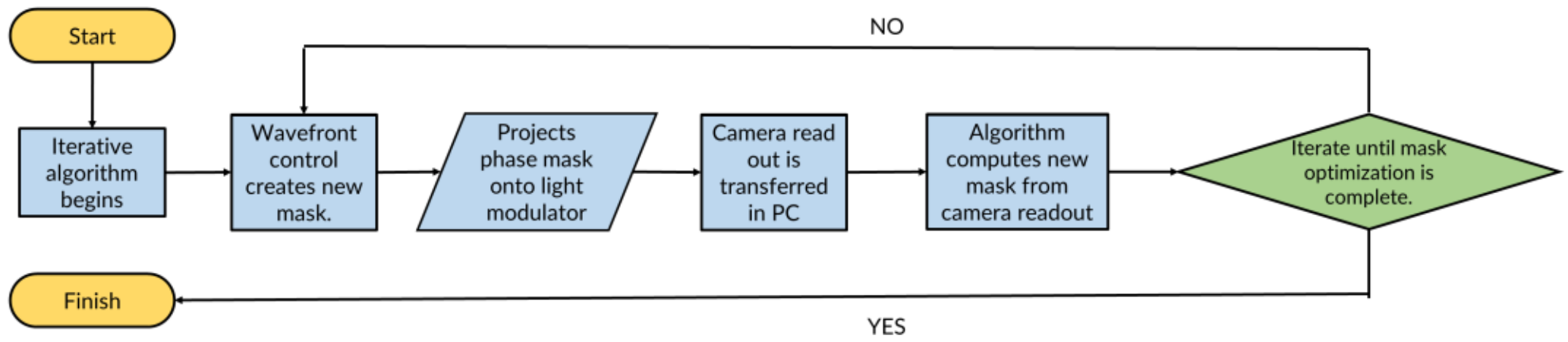

*Supplementary Figure 2. Illustration of the wavefront optimization process controlled by an iterative algorithm*

## Scattering performance

As a next step we have demonstrated the tolerance of Wavelens-based light sheet illumination to the sample's scattering properties, highlighting its enhanced performance over conventional optical elements. For this investigation, 1 mm thick gelatin slabs with varying $TiO_2$ concentrations have been employed as phantoms simulating soft tissue's scattering properties[28,29]. The slabs were inserted between the Wavelens device and the detection plane to record the optical intensity distribution, following the execution of the iterative step wise optimization algorithm to create a light-sheet structure behind the scattering slab. Similar measurements were also performed using the combined achromatic cylindrical lens and objective configuration for relative performance comparisons. In the case of optical scattering absence (slab without $TiO_2$ scattering factor), the generated light sheets were characterized by an apparently high homogeneity using either the conventional configuration or the Wavelens device (Supplementary Figure 1a). While the generation of the respective intensity profiles reveal quite smooth gaussian-like intensity distributions, the conventional configuration (red curve) incorporates the previously observed deviations from an ideal response, in the form of side-lobes, which can be mainly attributed to the presence of optical aberrations.

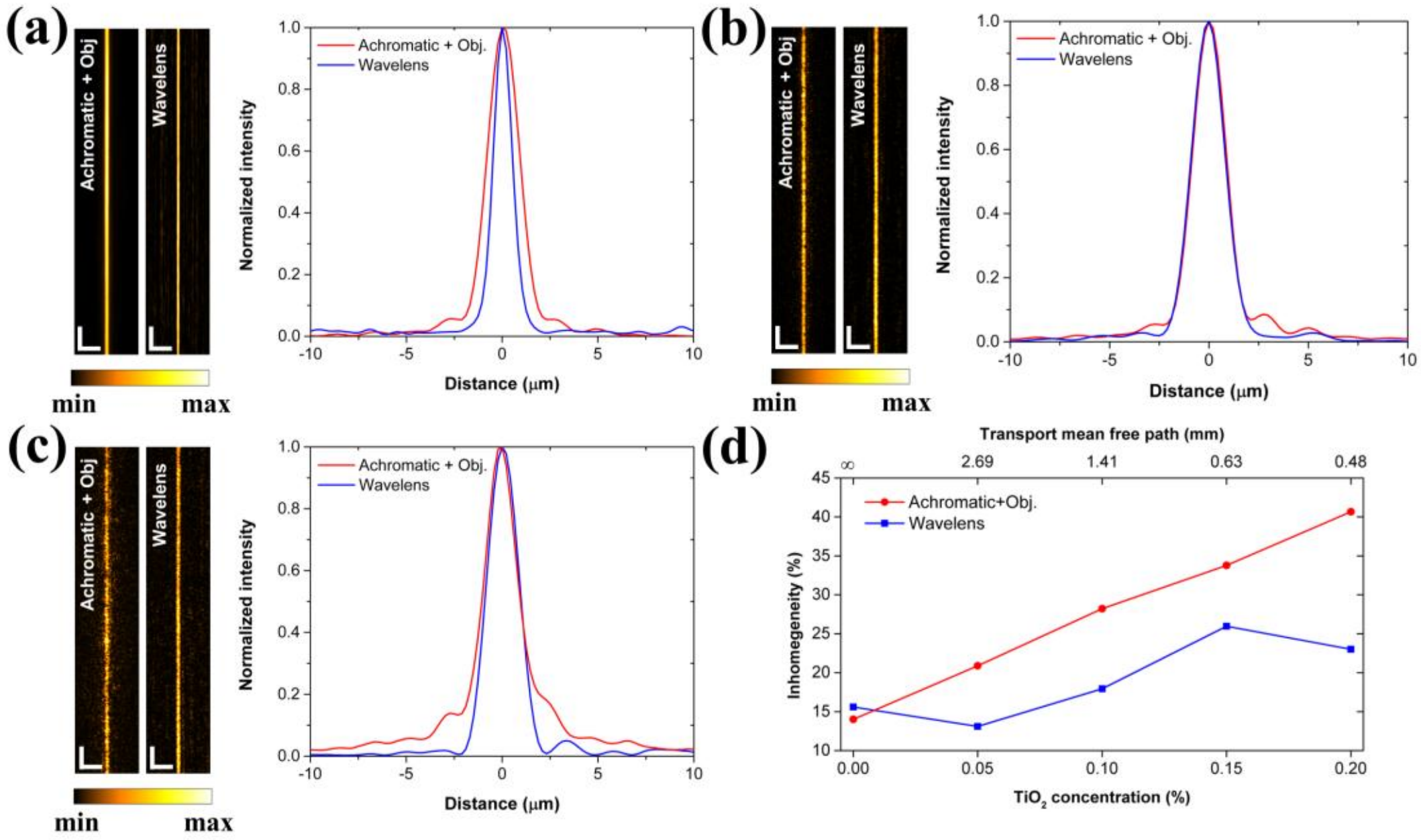

*Supplementary Figure 3. Performance comparison between Wavelens and conventional optics in scattering phantom samples.*

*(a) Light sheets produced by both conventional optics and Wavelens device in the absence of optical scattering (slab without TiO2). (b-c) Cross-sectional images and graphs of light sheets produced by both configuration in the presence of scattering ($TiO_2$ concentrations 0.1 and 0.2% respectively), showcasing the robustness of the intensity distribution by the Wavelens device, even under increased scattering. (Scale bar: 15μm). (d) Comparative graph of light sheet inhomogeneity as a function of $TiO_2$ concentration in the gelatin slab, demonstrating Wavelens superior robustness and preservation of homogeneity even under high scattering conditions, outperforming conventional optics by up to 45% for the slab containing 0.2% TiO2.*

By adding the $TiO_2$ medium at concentrations of 0.1 % (Supplementary Figure 3b) and 0.2% (Supplementary Figure 3c) respectively, the inhomogeneity of both light sheets appears to increase. Interestingly, the Wavelens, , despite the scattering induced by the slab, maintains a light sheet with well-defined intensity distribution (blue curves in Supplementary Figure 3b and 3c intensity profile plots), with significantly lower inhomogeneity. On the other hand, in the conventional optical configuration, a noticeable halo is evident around the central light distribution (red curves in Supplementary Figure 3b and 1c intensity profile plots). Finally, Supplementary Figure 3d presents a graph comparing the inhomogeneity of each light sheet produced by the conventional optics configuration (red line) and the Wavelens device (blue line) as a function of $TiO_2$ concentration in the gelatin slab (lower axis). The top axis of the graph corresponds to the measured transport mean free path in mm for each scattering slab using a commercial spectrophotometer (Perkin, Elmer Lambda 950 UV-Vis-NIR). In the case of conventional optics, an approximately linear relation between light sheet inhomogeneity and scattering strength can be observed. On the contrary, Wavelens device demonstrates substantially higher robustness as regards the effective preservation of light sheet homogeneity with increasing optical scattering, which can be at least 45% better for the slab containing 0.2% $TiO_2$.

The superior performance of Wavelens under high scattering conditions is linked to its inherent ability to manipulate optical wavefronts as they propagate through complex and random scattering media such as biological tissues, in contrast to the limited control provided by conventional elements.

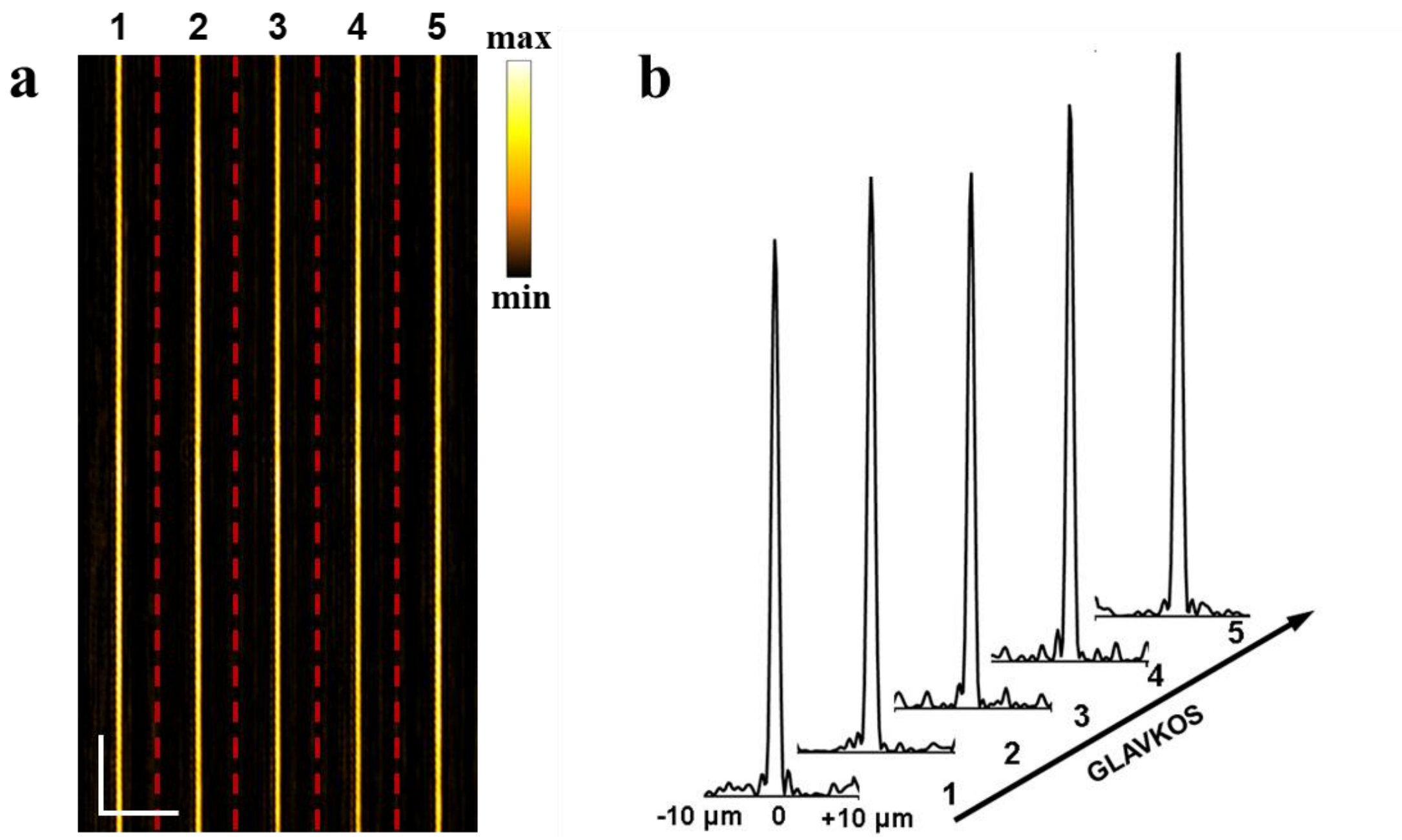

*Supplementary Figure 4: Optical Performance Assessment of the GLAVKOS.*

*(a) Intensity cross-sections at the focal plane of light sheets formed by five distinct GLAVKOS media at 488 nm illumination, showing high reproducibility. Scale bar:20μm. (b) Representative intensity profiles of the generated light sheets.*

Supplementary Figure 4a presents the resulting five distinct intensity cross-sections at the focal plane corresponding to the region where each light sheet has been formed, using an illumination wavelength at 488 nm. The normalized two-dimensional cross-correlation coefficients among the five images were calculated in the range of 0.977 ± 0.006 (mean value ± 1 standard deviation), demonstrating the high reproducibility of Wavelens response on the generation of uniform light sheet structures using different GLAVKOS scattering plates. Additionally, representative intensity profiles for the light sheets generated by utilizing five distinct GLAVKOS plates, are presented in Supplementary Figure 4a. The estimated average FWHM value was equal to 0.98 ± 0.05 μm, the respective inhomogeneity among the light sheets was 23.66 ± 2.31%, whereas the signal to noise ratio (SNR) range was 91.92 ± 9.71, further confirming the reliability and robustness of the proposed method.